\documentclass[twocolumn,times]{aastex631}

\usepackage{bm}
\usepackage{booktabs}
\usepackage{amsmath,amssymb}
\usepackage{multirow}
\begin{document}

\title{
Neural Networks as Surrogate Solvers for Time-Dependent Accretion Disk Dynamics}

\author[0009-0006-0889-3132]{Shunyuan Mao}
\affiliation{Department of Physics and Astronomy, Rice University \\
6100 Main St, Houston, TX 77005, USA, shunyuanm@gmail.com}
\affiliation{Department of Physics \& Astronomy, University of Victoria\\
3800 Finnerty Rd, Victoria, BC V8P 5C2, Canada}

\author[0000-0002-2798-8181]{Weiqi Wang}
\affiliation{Department of Physics \& Astronomy, University of Victoria\\
3800 Finnerty Rd, Victoria, BC V8P 5C2, Canada}

\author[0000-0002-7721-6884]{Sifan Wang}
\affiliation{Institute for Foundations of Data Science, Yale University \\
Kline Tower, 219 Prospect Street, New Haven, CT 06511, USA}

\author[0000-0001-9290-7846]{Ruobing Dong}
\affiliation{Kavli Institute for Astronomy and Astrophysics, Peking University \\
5 Yiheyuan Road, Haidian District, Beijing 100871, China, rbdong@pku.edu.cn}
\affiliation{Department of Physics \& Astronomy, University of Victoria\\
3800 Finnerty Rd, Victoria, BC V8P 5C2, Canada}

\author[0000-0002-5476-5768]{Lu Lu}
\affiliation{Department of Statistics and Data Science, Yale University \\
219 Prospect Street, New Haven, CT 06511, USA}

\author[0000-0001-9036-3822]{Kwang Moo Yi}
\affiliation{Department of Computer Science, University of British Columbia \\
2366 Main Mall, Vancouver, BC V6T 1Z4, Canada}

\author[0000-0002-2816-3229]{Paris Perdikaris}
\affiliation{Department of Mechanical Engineering and Applied Mechanics, University of Pennsylvania \\
220 S 33rd St, Philadelphia, PA 19104, USA}

\author[0000-0001-8061-2207]{Andrea Isella}
\affiliation{Department of Physics and Astronomy, Rice University \\
6100 Main St, Houston, TX 77005, USA}
\affiliation{Rice Space Institute,
Rice University \\
6100 Main St, Houston, TX 77005, USA}

\author[0000-0003-2239-7988]{Sébastien Fabbro}
\affiliation{National Research Council Canada \\
Herzberg Astronomy and Astrophysics Research Centre \\
5071 West Saanich Rd, Victoria, BC V9E 2E7, Canada}

\author[0000-0002-6540-7042]{Lile Wang}
\affiliation{Kavli Institute for Astronomy and Astrophysics, Peking University \\
5 Yiheyuan Road, Haidian District, Beijing 100871, China, rbdong@pku.edu.cn}

\begin{abstract}
Accretion disks are ubiquitous in astrophysics, appearing in diverse environments from planet-forming systems to X-ray binaries and active galactic nuclei. Traditionally, modeling their dynamics requires computationally intensive (magneto)hydrodynamic simulations. Recently, Physics-Informed Neural Networks (PINNs) have emerged as a promising alternative. This approach trains neural networks directly on physical laws without requiring data.  
We for the first time demonstrate PINNs for solving the two-dimensional, time-dependent hydrodynamics of non-self-gravitating accretion disks.
Our models provide solutions at arbitrary times and locations within the training domain, and successfully reproduce key physical phenomena, including the excitation and propagation of spiral density waves and gap formation from disk-companion interactions. Notably, the boundary-free approach enabled by PINNs naturally eliminates the spurious wave reflections at disk edges, which are challenging to suppress in numerical simulations.
These results highlight how advanced machine learning techniques can enable physics-driven, data-free modeling of complex astrophysical systems, potentially offering an alternative to traditional numerical simulations in the future. 
\end{abstract}


\section{Introduction} \label{sec:intro}
Accretion disks are governed by nonlinear, time-dependent partial differential equations (PDEs), such as the Navier–Stokes equations, and exhibit complex, multiscale dynamics \citep{frank2002accretion, armitage2011dynamics}.  
Traditional numerical methods, such as grid-based hydrodynamic solvers FARGO3D \citep{masset2000fargo,benitez2016fargo3d} and Athena++ \citep{stone2020athena++}, excel at capturing these dynamics. However, such simulations often demand immense computational resources, sometimes requiring millions of CPU hours for a single run \citep{huang2025dust}.
The computational burden grows further when exploring large parameter spaces to fit observations with increasingly rich detail. 

For instance, the superb resolution and sensitivity of the Atacama Large Millimeter/submillimeter Array have provided an unprecedented view of substructures in protoplanetary disks \citep[e.g.,][]{alma2015hl}, enabling us to infer forming planets from observations that are otherwise difficult to detect \citep{paardekooper2023planet}. 
Reproducing the morphology of individual disks with companion–disk interaction simulations, however, remains labor-intensive \citep{dipierro2015planet,dong2015observational2,jin2016modeling}.  
While substructures have been characterized in hundreds of disks 
\citep{andrews2020observations,bae2023structured}, embedded companion candidates have been inferred in only a few dozen \citep{zhang2018disk,lodato2019newborn}.
Developing faster and more flexible modeling techniques is therefore increasingly urgent to keep pace with these observational advances.

Recent advances in machine learning (ML) offer promising alternatives for solving PDEs. 
Many of these new techniques are data-driven \citep{li2020fourier,kovachki2023neural,mccabe2023multiple,herde2024poseidon}, training neural networks on large datasets of pre-computed PDE solutions to act as fast surrogate models. While often achieving significant speedups, the fundamental limitation of this approach is its reliance on these large training datasets, which are typically unavailable or prohibitively expensive to generate for many astrophysical problems.

In contrast, Physics-Informed Neural Networks \citep[PINNs,][]{raissi2019physics} have opened new avenues for scientific computing. PINNs stand out as they do not require solution data to train; instead, they learn the solution by embedding the governing equations directly into the loss function, enabling data-free learning that respects physical laws. PINNs have shown success in relatively simple systems, such as fluid dynamics and solid mechanics \citep{cai2021physics,faroughi2024physics}, and hold potential for astrophysical fluid dynamics applications \citep{auddy2024grinn}. Yet, their ability to tackle challenging systems, including those with long-term evolution, multi-scale dynamics, and sharp discontinuities \citep{karniadakis2021physics,wang2023expert,zhang2024physics}, remains an open question. 

To address the question of whether PINNs can solve the time-dependent Navier–Stokes equations in accretion disks without pre-existing data and offer a novel alternative to traditional simulations, we apply PINNs to solve the two-dimensional (2D) hydrodynamics of an accretion disk perturbed by a low mass companion (\S\ref{sec:math_model}). 
As a direct application of PINNs is insufficient, we introduce several techniques to overcome the limitations of the standard approach: 
(i) a time-marching scheme that decomposes the long-term evolution into a sequence of shorter, manageable time intervals (\S\ref{sec:time_marching}); 
(ii) a radial boundary strategy to suppress spurious reflections near domain edges (\S\ref{sec:radial_boundary_conditions}); 
(iii) a periodic coordinate transform to enforce azimuthal periodicity (\S\ref{sec:azimuthal_boundary_conditions}); 
(iv) a strategy to enforce the initial conditions, preventing the accumulation of small errors that lead to incorrect solutions over time (\S\ref{sec:initial_condition_and_hard_constraint}); 
(v) an output-scaling strategy to avoid poor network initialization (\S\ref{sec:initialization_bias});
and 
(vi) a loss-balancing strategy for the competing loss terms (\S\ref{sec:competing_loss_terms_and_weight_balancing}).
We validate our method against several FARGO3D simulations with varied parameters (\S\ref{sec:tests}), and discuss the capabilities and limitations of our approach (\S\ref{sec:conclusion}).

\section{Physics-informed Neural Network implementation} \label{sec:physics_informed_training}

\begin{figure*}
    \includegraphics[width=\textwidth]{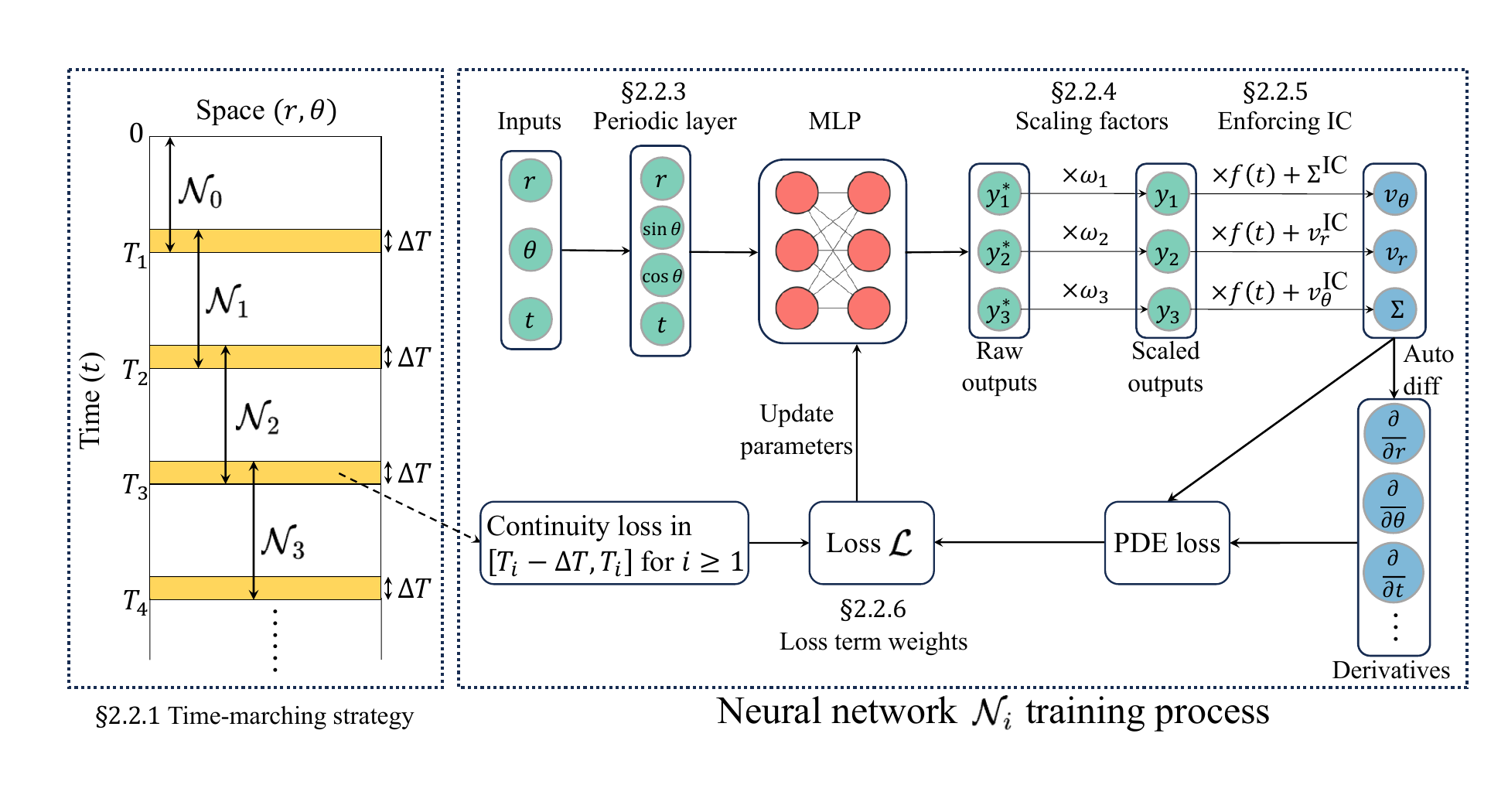}
    \caption{
    The training workflow for our PINN. The overall strategy (left) partitions the total simulation time $[0,T]$ into a sequence of overlapping temporal windows (shown in yellow), such as $[0,\ T_1], [T_1-\Delta T,\ T_2], [T_2-\Delta T,\ T_3]$, etc., each with its own dedicated neural network $\mathcal{N}_0$, $\mathcal{N}_1$, $\mathcal{N}_2$, etc. (\S\ref{sec:time_marching}). The procedure for updating the network within a single window is detailed on the right.
    The process begins by feeding spatio-temporal coordinates $(r, \theta, t)$ into a Multi-Layer Perceptron (MLP). A periodic layer then acts on the angular coordinate $\theta$ to enforce the required azimuthal boundary conditions (\S\ref{sec:azimuthal_boundary_conditions}). The raw MLP outputs are subsequently scaled by factors, $\omega_i$, and combined with initial profiles to construct the final solution variables $(\Sigma, v_r, v_\theta)$ (\S\ref{sec:initialization_bias}, \S\ref{sec:initial_condition_and_hard_constraint}). These variables are then processed via automatic differentiation to compute the residuals of the governing PDEs, which form the PDE loss component (\S\ref{sec:competing_loss_terms_and_weight_balancing}). The total loss function, $\mathcal{L}$, combines this PDE loss with a continuity loss. The second loss ensures the network $\mathcal{N}_i$ gives the same solution as $\mathcal{N}_{i-1}$ inside the overlapping region $[T_i-\Delta T,\ T_i]$ for $i \geq 1$. Finally, a gradient-based optimizer iteratively updates the network's parameters to minimize the total loss $\mathcal{L}$.}
    \label{fig:PINN_train} 
\end{figure*}

We design and train PINNs to map spatial and temporal coordinates $\text{(radius, azimuth, time)} = (r,\theta,t)$ to the physical solution variables: surface density $\Sigma$, radial velocity $v_r$, and azimuthal velocity $v_\theta$. The workflow of the overall training process is illustrated in Figure \ref{fig:PINN_train}, and the hyperparameters of the network are listed in Table \ref{tab:ml_hyperparams}. We introduce the vanilla PINN approach in \S\ref{sec:vanilla_pinn}, then key modifications in \S\ref{sec:pinn_modification}. 

\subsection{Introduction to the vanilla PINN}\label{sec:vanilla_pinn}
PINNs are a machine learning approach to solve differential equations by incorporating the equations directly into the loss function of a neural network, which typically uses a Multi-Layer Perceptron (MLP) as the function approximator \citep{pinkus1999approximation}.
A key feature is that the neural network, as a continuous function, provides solutions at arbitrary times and locations, offering a practical advantage over traditional solvers, which can only provide solutions on discretized grids.
In addition to the equations, the loss function may also contain data-driven components to ensure that the solution satisfies the initial and boundary conditions. The method has been recently reviewed by \citet{karniadakis2021physics}.

The physics-informed loss is computed at a set of collocation points sampled within the spatio-temporal domain. These points are locations where the network's output is evaluated against the governing physical laws. At each collocation point, automatic differentiation is employed to compute the necessary derivatives of the network's outputs (e.g., $\Sigma$, $v_r$, and $v_\theta$ in our case) with respect to its inputs (e.g., $r,\theta,t$), allowing for the direct evaluation of the PDE residuals. The network's parameters are then iteratively updated, typically using a gradient-based optimizer such as Adaptive Moment Estimation \citep[Adam,][]{kingma2014adam}, to drive the total loss toward zero. This process encourages the network's outputs to approximate a solution that honors the physical laws encoded in the PDEs across the entire domain.

While the `vanilla' PINN formulation has shown promise for many problems \citep{raissi2019physics,mao2020physics,jin2021nsfnets}, its application to complex systems in astrophysical fluid dynamics, such as accretion disks, faces substantial challenges. In the remaining of \S\ref{sec:physics_informed_training}, we discuss the limitations of this approach and the modifications required for our application.

\begin{table*}
    \centering
    \begin{tabular}{ l l c }
        \toprule
        \toprule
        \textbf{Category} & \textbf{Hyperparameter} & \textbf{Default Value} \\
        \midrule
        \midrule
        \multirow{3}{*}{Architecture}
            & Layer size & 32 neurons per layer $\times$ 9 hidden layers \\
            & Initialization & Glorot (Xavier) normal~\citep{glorot2010understanding} \\
            & Activation function & Stan~\citep{gnanasambandam2022self} \\
        \midrule
        \multirow{8}{*}{Training}
            & Learning Rate (LR) & 0.001 \\
            & LR decay rate & 0.9 \\
            & LR transition steps & 5000 \\
            & Sample size & 10,000 \\
            & Sampling distribution & Uniform random \\
            & Number of training steps & 32 $\times$ 200,000 $=$ 6,400,000 \\
            & Optimizer & Adam~\citep{kingma2014adam} \\
            & Number of windows & 32 \\
            & Weight balancing method & Initial loss normalization\\
        \midrule
        \multirow{3}{*}{Scaling}
            & Scaling factor $\omega_1$ & $0.1$ ($M_{\rm 2}/M_{\rm 1}=0.001$) or $1$ ($M_{\rm 2}/M_{\rm 1}=0.0001$) \\
            & Scaling factor $\omega_2$ & $0.001$ ($M_{\rm 2}/M_{\rm 1}=0.001$) or $0.05$ ($M_{\rm 2}/M_{\rm 1}=0.0001$) \\ 
            & Scaling factor $\omega_3$ & $0.001$ ($M_{\rm 2}/M_{\rm 1}=0.001$) or $0.05$ ($M_{\rm 2}/M_{\rm 1}=0.0001$) \\
        \bottomrule
    \end{tabular}
    \caption{
    Hyperparameters for the PINN models presented in \S\ref{sec:tests}, including network architecture, training parameters, and scaling factors.
    The network architecture is a Multi-Layer Perceptron (MLP), with its structure specified in the `Layer size' row.
    The scaling factors are tuned for different values of $M_{\rm 2}/M_{\rm 1}$ to ensure the network output is of the same order of magnitude as the target solution. Except for the scaling factors, all other hyperparameters are held constant for the models in \S\ref{sec:tests}.}
    \label{tab:ml_hyperparams}
\end{table*}

\subsection{
Modifications to the vanilla PINNs and training}\label{sec:pinn_modification}
We implement a series of modifications on top of the vanilla PINN to improve its performance and physical accuracy. To manage training complexity, we adopt a time-marching technique that breaks the problem into smaller, sequential steps (\S\ref{sec:time_marching}). We enforce azimuthal periodicity architecturally and intentionally avoid explicit radial boundary conditions to ensure the solution exhibits the expected wave propagation behavior (\S\ref{sec:radial_boundary_conditions} and \S\ref{sec:azimuthal_boundary_conditions}). Furthermore, we introduce output scaling factors to mitigate initialization biases (\S\ref{sec:initialization_bias}). The system's initial state is enforced using a hard constraint method, which guarantees that the network's predictions satisfy the initial conditions without requiring training (\S\ref{sec:initial_condition_and_hard_constraint}). Finally, we balance competing loss components to ensure the network learns all physical laws simultaneously (\S\ref{sec:competing_loss_terms_and_weight_balancing}).

\subsubsection{Long-term integration and time-marching technique}\label{sec:time_marching}
A single PINN struggles to learn PDE solutions across an entire spatio-temporal domain, particularly in long-term simulations \citep{wang2023expert,wang2024respecting}. In problems like modeling companion-disk interactions, the network must capture complex dynamics over vast timescales, including distinct behaviors at different stages, e.g., density wave excitation and gap opening. Such a task demands immense network capacity and can lead to training failures. To overcome this challenge, we implement a time-marching strategy that decomposes the simulation into a sequence of smaller, more manageable time intervals \citep{wight2020solving}.

As illustrated in Figure \ref{fig:PINN_train}, this strategy partitions the total simulation time $T$ into a sequence of overlapping temporal windows (shown in yellow), $[0,\ T_1]$, $[T_1-\Delta T,\ T_2]$, $[T_2-\Delta T,\ T_3]$, etc., with a separate, small neural network trained for each window. This approach simplifies the learning task for each network, as it only needs to approximate the solution's local behavior within a short time interval instead of capturing the dynamics over the entire simulation.

The networks are trained sequentially, starting with the initial time window. To ensure a continuous solution, the final state from a fully trained network in one window provides the initial conditions for the network in the subsequent window. Specifically, the surface density and velocities at the end of one window serve as labeled training data for the next, with continuity enforced within the overlapping region (see Figure~\ref{fig:PINN_train}). The subsequent network is then trained using a combined loss from both this labeled data and the PDE residuals in its domain. Because the network in each window takes time $t$ as a continuous input, it can predict the solution at any time within that window after training. By linking a series of networks in this manner, we model the system's evolution from its initial state to the final time $T$. Test results suggest that the time-marching approach is more effective than a single network for learning the entire process.

\subsubsection{Radial boundary conditions and wave-killing technique}\label{sec:radial_boundary_conditions}
Traditional numerical methods using techniques such as finite elements and finite volume require explicit boundary conditions at the inner and outer radial boundaries, $r_{\rm min}$ and $r_{\rm max}$, to solve the system. To handle wave reflection, these methods often need to employ wave-killing mechanisms, such as establishing damping zones \citep{de2006comparative,de2007vortex,edelmann2019three,peric2020reducing}. While a vanilla PINN could incorporate such boundary constraints in its loss function, this additional penalty would likely compete with the primary PDE loss component and hinder the training convergence.

In contrast, our boundary-free approach eliminates the need for explicit boundary conditions. Unlike traditional methods, we neither introduce specialized architectural components nor incorporate radial boundary conditions into the loss function \citep{rasht2022physics,wang2023acoustic,wang2023physics,ding2023self}. We find that our network naturally converges to a solution that suppresses waves at the boundaries without any explicit constraints.

This emergent wave-killing behavior is a consequence of the Multi-Layer Perceptron (MLP) architecture's inherent smoothness bias. Constructed with continuous and differentiable activation functions, MLPs naturally favor smooth function approximations and extrapolation, a finding consistent with previous work \citep{rasht2022physics}. Consequently, when the network is trained solely through PDE residual minimization, this smoothness bias guides the network to learn a solution that satisfies the governing physics inside the domain while smoothly extrapolating at the radial boundaries, thereby suppressing spurious reflections. This approach not only avoids gradient competition from boundary loss components but also eliminates potential errors from artificial boundary treatments, allowing the system to evolve more naturally, especially where waves interact with the domain edges.

\subsubsection{Azimuthal boundary conditions and periodic coordinate transformation}\label{sec:azimuthal_boundary_conditions}
Accretion disks are periodic with respect to the azimuthal coordinate, which requires that physical quantities ($\Sigma$, $v_r$, $v_\theta$) and their partial derivatives be identical at $\theta = -\pi$ and $\theta = \pi$. A traditional PINN approach enforces this requirement by adding explicit periodic boundary conditions to the loss function. This strategy, however, complicates training, as the optimizer must balance these boundary constraints with the PDE residuals. Furthermore, this loss-based enforcement means periodicity is only approximately satisfied until convergence, which can create a discontinuity at the azimuthal boundary.

To address these challenges, we apply a coordinate transformation that enforces periodicity. By converting the input coordinates from $(r,\theta,t)$ to $(r,\sin\theta,\cos\theta,t)$, as shown in Figure~\ref{fig:PINN_train}, the network input becomes inherently periodic. This transformation ensures the predicted solution is automatically continuous at the $\theta = \pm\pi$ boundary, eliminating the need for an explicit boundary loss term. We choose this representation over $(r \sin\theta,r \cos\theta,t)$ because it provides a more effective basis for the network to model features such as spiral arms and gaps in the solution.

\subsubsection{Network initialization bias and output scaling}\label{sec:initialization_bias}
While standard initialization schemes like Glorot (Xavier) \citep{glorot2010understanding} configure network parameters to produce an output variance near $1.0$, the variance of the true solution (e.g., for surface density) can be orders of magnitude different. This initial discrepancy in scale forces the optimizer to aggressively adjust parameters across all layers merely to correct the output's magnitude. This preliminary, corrective phase can prolong convergence, increase the risk of entrapment in suboptimal local minima, and render the training process highly sensitive to the chosen learning rate.

We mitigate this initialization bias by applying a set of non-trainable scaling factors, $\omega_1$, $\omega_2$, and $\omega_3$, to the network's final output (Figure~\ref{fig:PINN_train}). Instead of being learned, these factors serve as hyperparameters manually tuned with prior knowledge of the true solution's magnitude (Table~\ref{tab:ml_hyperparams}). This method bridges the magnitude gap between the network's initial output and the target solution, thereby enhancing both training efficiency and convergence stability.

\subsubsection{Initial conditions and hard constraints}\label{sec:initial_condition_and_hard_constraint}
For time-dependent PDEs, accurate enforcement of the initial condition is critical to the accuracy of the predicted solution. Standard PINNs include the initial condition as a loss component in the total loss function, which means the initial condition error is gradually minimized during training. However, any small errors in the initial condition can propagate to later times and become amplified, eventually causing the network to learn an incorrect solution.

To enforce the initial conditions by construction, we formulate the network output to be zero at $t=0$ and then add the known initial condition function \citep{mcfall2009artificial}. Combined with the output scaling factors (\S\ref{sec:initialization_bias}), the network output transformation is:
\begin{equation}
\begin{array}{rcl}
    \Sigma^{\text{Predict}}(r, \theta, t) &=& \Sigma^{\text{IC}}(r, \theta) + f(t) \omega_{\rm 1} y_{\rm 1}^{\star}(r, \theta, t), \\
    v_r^{\text{Predict}}(r, \theta, t) &=& v_r^{\text{IC}}(r, \theta) + f(t) \omega_{\rm 2} y_{\rm 2}^{\star}(r, \theta, t), \\
    v_{\theta}^{\text{Predict}}(r, \theta, t) &=& v_{\theta}^{\text{IC}}(r, \theta) + f(t) \omega_{\rm 3} y_{\rm 3}^{\star}(r, \theta, t),
\end{array}
\end{equation}
where $f(t) = 1 - e^{-t/T_1}$ for the first window $t\in [0, T_1]$ and $f(t)=1$ for subsequent windows. The functions $\Sigma^{\text{IC}}(r, \theta)$, $v_r^{\text{IC}}(r, \theta)$ and $v_{\theta}^{\text{IC}}(r, \theta)$ represent the initial conditions. The terms $y_{\rm 1}^{\star}$, $y_{\rm 2}^{\star}$, and $y_{\rm 3}^{\star}$ are the outputs of the network's final layer.

\subsubsection{Competing loss components and weight balancing}\label{sec:competing_loss_terms_and_weight_balancing}
Our method for handling boundary conditions (\S\ref{sec:radial_boundary_conditions} and \S\ref{sec:azimuthal_boundary_conditions}) and initial conditions (\S\ref{sec:initial_condition_and_hard_constraint}) eliminates these components from the loss function, thereby avoiding competition with the PDE residual components. However, balancing the loss function contributions from different PDEs and time-marching continuity data remains a challenge. Improper balancing can lead to poor convergence or inaccurate solutions, especially since the relative magnitudes of these loss components can vary significantly across the domain. This necessitates a careful loss weighting strategy.

We address this challenge using fixed weights to balance the different scales of the loss components. These weights are determined from the loss values computed at initialization, which provides a good estimate of each component's scale. The reweighting scheme mitigates scale differences among the loss components and ensures their contributions to the optimization process are balanced.

\section{Accretion Disk Model} \label{sec:math_model}
We model the time-dependent evolution of a 2D accretion disk perturbed by a low-mass companion (secondary) on a fixed circular orbit around a central object (primary). To simplify the problem, we transform the governing equations into a non-inertial reference frame centered on the primary and rotating with the secondary's constant angular velocity, $\Omega = \sqrt{G(M_{\rm 1} + M_{\rm 2})/r_{\rm 2}^3}$, where $G$ is the gravitational constant, $M_{\rm 1}$ and $M_{\rm 2}$ are the primary and secondary's masses, respectively, and $r_{\rm 2}$ is the secondary's orbital radius. In this frame, the secondary remains fixed at the coordinates $r=r_{\rm 2}$ and $\theta=0$. The total gravitational potential, $\phi$, in the rotating frame includes contributions from the primary, the secondary, and an indirect term accounting for the frame's acceleration:
\begin{eqnarray} \label{eq:gravity_potential}
\phi(r,\theta) &=& \phi_{\rm 1} + \phi_{\rm 2} + \phi_{\text{ind}} \\
&=& -G\frac{M_{\rm 1}}{r} - G\frac{M_{\rm 2}}{\sqrt{r^2 + r_{\rm 2}^2- 2r r_{\rm 2}\cos{\theta}+s^2}} \nonumber \\
& & + G\frac{M_{\rm 2} r \cos{\theta}}{r_{\rm 2}^2},
\end{eqnarray}
where a smoothing length, $s = 0.6 H_{\rm 2}$, is introduced into the secondary's potential, $\phi_{\rm 2}$, to prevent a numerical divergence at its location; $H_{\rm 2}$ is the disk scale height at the secondary's orbital radius. The indirect term, $\phi_{\text{ind}}$, arises because the origin of our coordinate system (the primary) accelerates due to the gravitational pull of the secondary.

In this reference frame, the mass conservation and momentum equations are:
\begin{align}
    &\frac{\partial \Sigma}{\partial t} + \bm{\nabla} \cdot(\Sigma\bm{v}) = 0, \label{eq:PDE1_inertial} \\
    &\frac{\partial(\Sigma v_r)}{\partial t} + \bm{\nabla}\cdot(\Sigma v_r \bm{v}) - \Sigma \left(\frac{(v_{\theta}+r\Omega)^2}{r}-\frac{G M_{\rm 1}}{r^2}\right) \nonumber\\
    &\quad + \frac{\partial p}{\partial r} + \Sigma\frac{\partial \phi_{\rm 2}}{\partial r} + \Sigma\frac{\partial \phi_{\text{ind}}}{\partial r} - f_{r} = 0, \label{eq:PDE2_rotating} \\
    &\Sigma\frac{\partial L}{\partial t} + \bm{\nabla} L\cdot \left(\Sigma \bm{v}\right) \nonumber\\
    &\quad + \frac{\partial p}{\partial \theta} + \Sigma \frac{\partial \phi_{\rm 2}}{\partial \theta} + \Sigma \frac{\partial \phi_{\text{ind}}}{\partial \theta} - r f_{\theta} = 0, \label{eq:PDE3_rotating}
    \end{align}
where $\Sigma$ is the two-dimensional gas surface density, $\bm{v}$ is the velocity vector in polar coordinates ($r$, $\theta$) with radial and azimuthal components $v_r$ and $v_\theta$; 
$p$ is the vertically integrated pressure, and assuming the disk is locally isothermal, is given by $p = c_s^2 \Sigma$, where the time-invariant sound speed is $c_s=h v_K$, with $v_K= \sqrt{G M_{\rm 1}/r}$ being the Keplerian velocity and $h$ a constant disk aspect ratio; 
$L=r (v_{\theta}+r\Omega)$ is the angular momentum per unit mass in the inertial frame;
and $f_r$ and $f_\theta$ are the components of the viscous force per unit density, given by the divergence of the viscous stress tensor $\bm{T}$. For 2D flow, $\bm{T}$ is given by:
\begin{equation}
\bm{T}= \mu \left[ \bm{\nabla} \bm{v} + (\bm{\nabla} \bm{v})^{\intercal} - \frac{2}{3} (\bm{\nabla} \cdot \bm{v}) \bm{I} \right],
\end{equation}
where $\bm{I}$ is the identity tensor and $\mu = \Sigma \nu$ is the dynamic viscosity.  The kinematic viscosity, $\nu$, is constant and set to $10^{-5}$ in code units.
We initialize a smooth, axisymmetric disk in approximate equilibrium before the secondary sets in: $\Sigma(t=0)=\Sigma_0$ (constant), $v_r(t=0)=0$, $v_\theta(t=0)=\sqrt{ (1-h^2)G M_{\rm 1}/r } - r\Omega$, the sub-Keplerian velocity under the radial pressure gradient.

To generate reference solutions for the validation tests (\S\ref{sec:tests}), we run FARGO3D simulations. The computational domain extends radially from $r_{\text{min}}=0.4r_{\rm 2}$ to $r_{\text{max}}=2.5r_{\rm 2}$ with a grid resolution of $1024\times1536$ cells in the radial and azimuthal directions, respectively. Each simulation runs for 10 orbits. For the fiducial case, we set the mass ratio to $M_{\rm 2}/M_{\rm 1}=0.001$ and the disk aspect ratio to $h=0.05$ (\S\ref{sec:solutions_for_q1e-3}). We then test two additional scenarios in \S\ref{sec:performance_under_different_disk_parameters}: a low secondary mass case ($M_{\rm 2}/M_{\rm 1}=0.0001$, $h=0.05$) and a high disk aspect ratio case ($M_{\rm 2}/M_{\rm 1}=0.001$, $h=0.1$).

For the radial boundaries, we adopt the default open boundary conditions in FARGO3D.
These conditions enforce $v_r=0$ at the boundary, extrapolate the surface density ($\Sigma$) and azimuthal velocity ($v_\theta$), and employ damping zones to minimize wave reflection. We apply the periodic boundary conditions in the azimuthal direction.

\section{Tests}\label{sec:tests}
We train our PINN model using the hyperparameters listed in Table \ref{tab:ml_hyperparams}. The model results are validated against FARGO3D simulations, first for a fiducial case (\S\ref{sec:solutions_for_q1e-3}) and then for cases with varying parameters (\S\ref{sec:performance_under_different_disk_parameters}). The effect of the non-reflecting radial boundaries in the PINN is also highlighted in \S\ref{sec:performance_under_different_disk_parameters}.

\subsection{A fiducial case}
\label{sec:solutions_for_q1e-3}
\begin{figure*}
    \centering
    \includegraphics[width=0.8\linewidth]{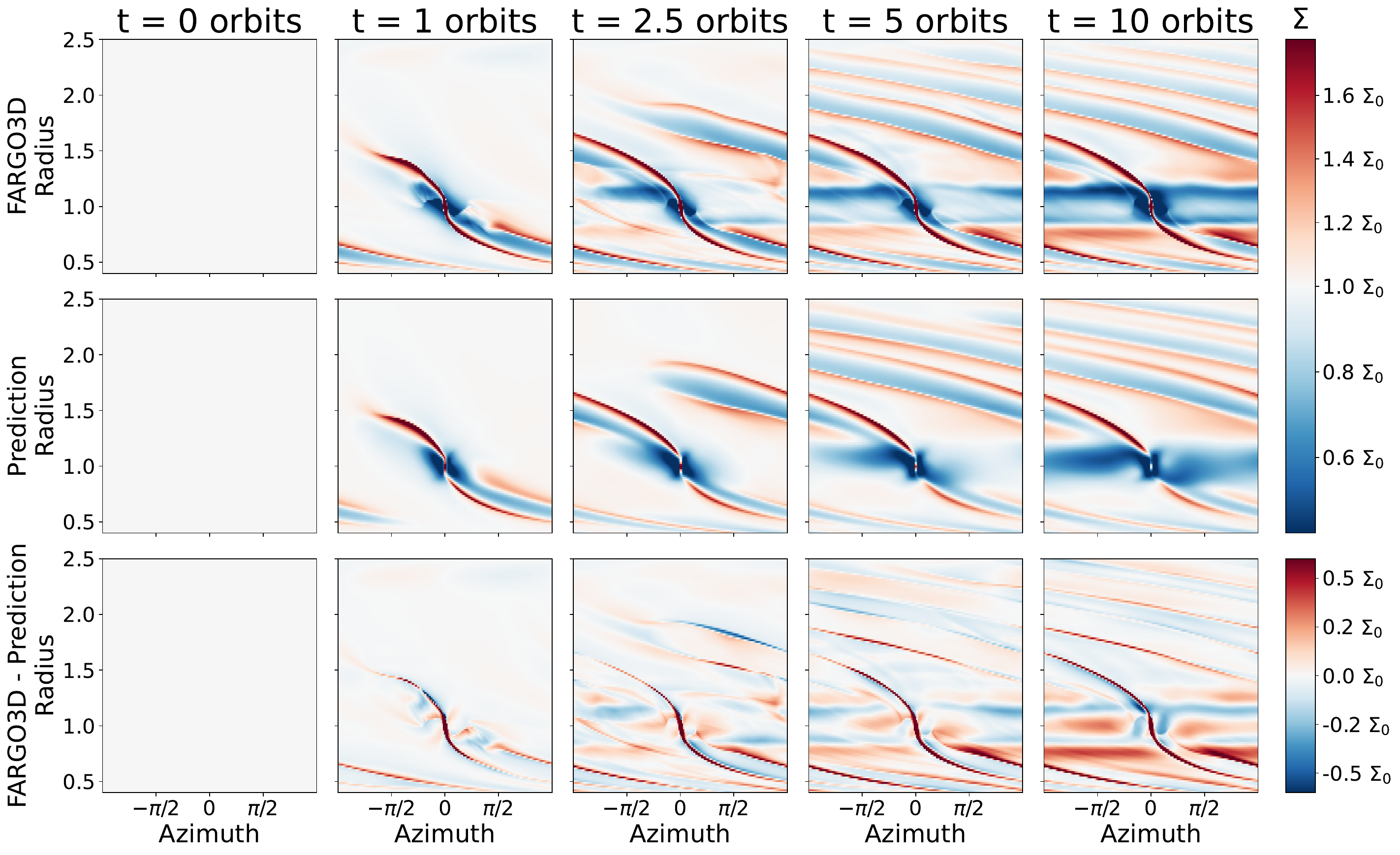}
    \caption{
    Comparison of surface density ($\Sigma$) predicted by the PINN (middle row) and simulated by FARGO3D (top row) for the fiducial case with a mass ratio $M_{\rm 2}/M_{\rm 1} = 10^{-3}$. The bottom row shows the difference between FARGO3D and PINN predictions (FARGO3D minus PINN), with a different color scale than the top and middle rows to better highlight the discrepancies. From left to right, the columns show snapshots at $0$, $1$, $2.5$, $5$ and $10$ orbits. An animation of this figure is available at \url{https://doi.org/10.6084/m9.figshare.30192904}.
}
    \label{fig:pred_fargo_compare_sigma}
\end{figure*}

\begin{figure*}
    \centering
    \includegraphics[width=0.8\linewidth]{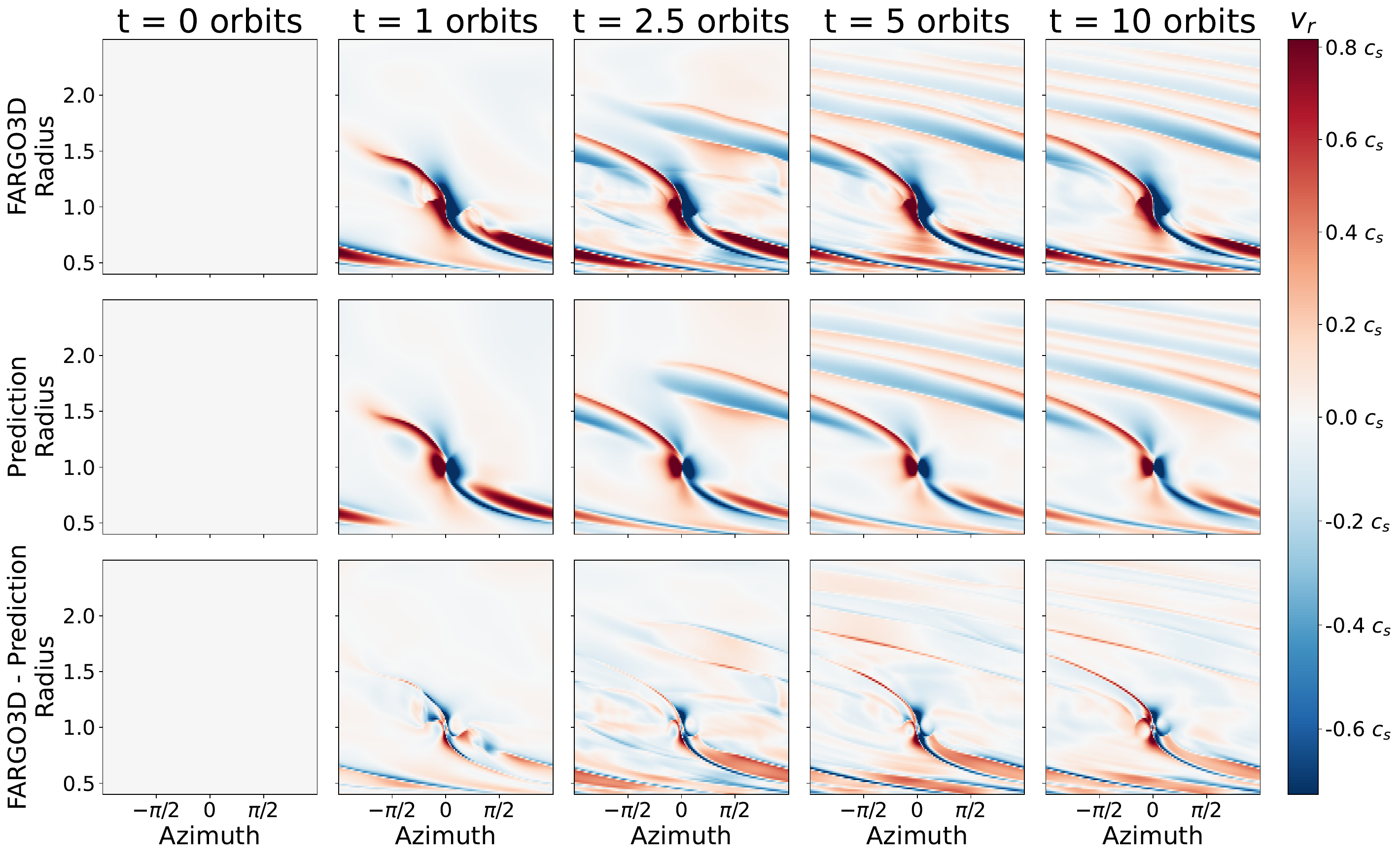}
    \caption{Same as Figure~\ref{fig:pred_fargo_compare_sigma}, but for the radial velocity $v_r$. An animation of this figure is available at \url{https://doi.org/10.6084/m9.figshare.30192904}.}
    \label{fig:pred_fargo_compare_vr}
\end{figure*}

\begin{figure*}
    \centering
    \includegraphics[width=0.8\linewidth]{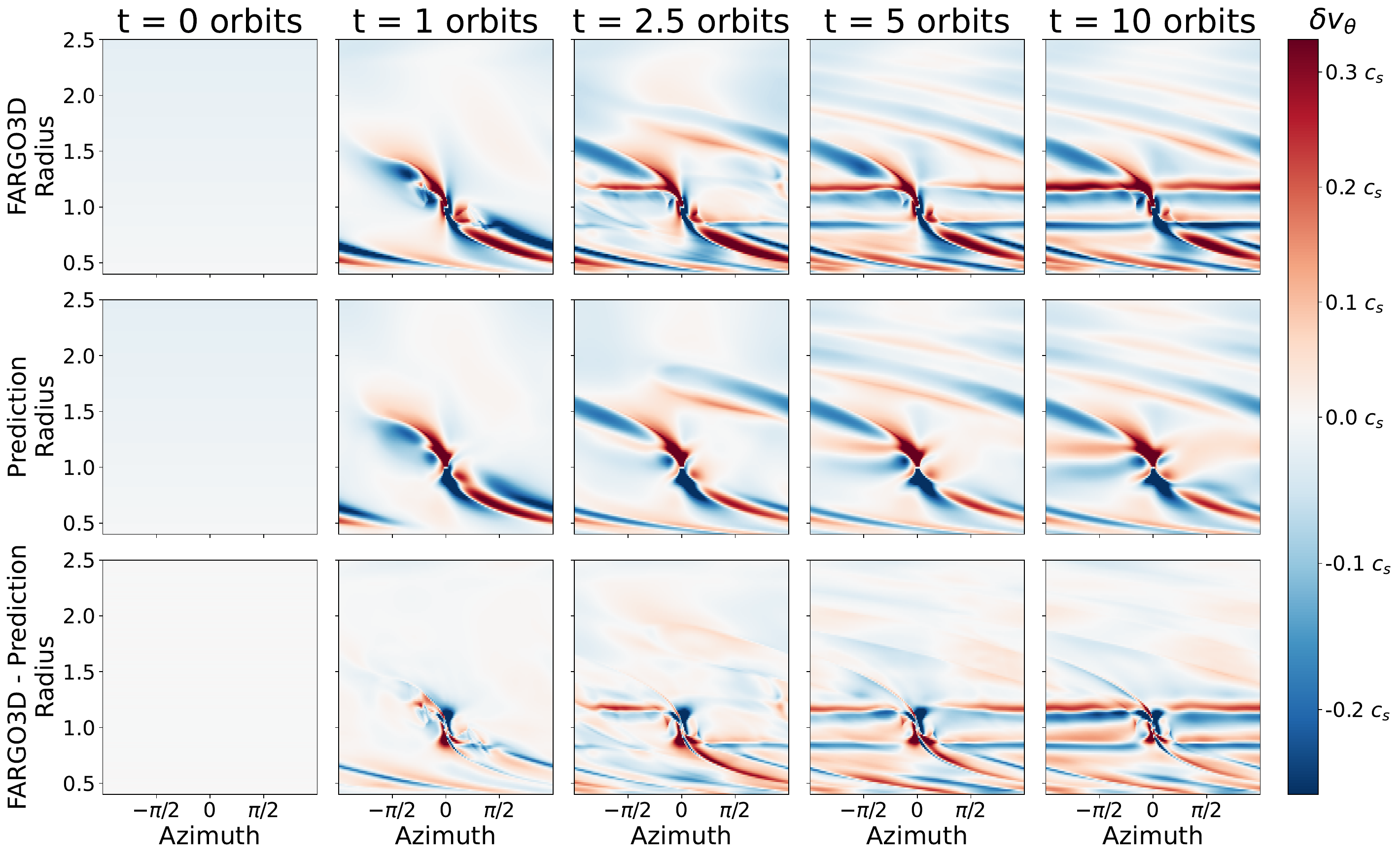}
    \caption{
    Same as Figure~\ref{fig:pred_fargo_compare_sigma}, but for the $\delta v_\theta(t) = v_\theta(t) - v_\theta(t=0)$. An animation of this figure is available at \url{https://doi.org/10.6084/m9.figshare.30192904}.
    }
    \label{fig:pred_fargo_compare_vtheta}
\end{figure*}

We first examine a fiducial case with a mass ratio of $M_{\rm 2}/M_{\rm 1}=10^{-3}$ and a disk aspect ratio of $h=0.05$. In Figures \ref{fig:pred_fargo_compare_sigma}, \ref{fig:pred_fargo_compare_vr}, and \ref{fig:pred_fargo_compare_vtheta}, we compare the density, radial velocity, and azimuthal velocity in polar coordinates from the PINN prediction with those from FARGO3D simulations. To highlight the perturbation generated by the secondary, we subtract the initial profile $v_\theta(t=0)=\sqrt{ (1-h^2)G M_{\rm 1}/r } - r\Omega$ from the azimuthal velocity field and plot the resulting difference in Figure \ref{fig:pred_fargo_compare_vtheta}. We examine the PINN’s predictions for the spiral arms and the gap. The secondary excites two prominent trailing arms that propagate through the disk, one inward and one outward. After 10 orbits, a low–surface-density annular region, or gap, forms around the secondary’s orbit. The largest discrepancies between the two methods occur at the sharp edges of the gap and spiral arms, where the PINN’s features appear smoother than those in the FARGO3D reference. Nevertheless, the PINN accurately reproduces the overall morphology and captures the substantial density depletion within the gap.


\begin{figure*}
    \centering
    \includegraphics[width=0.8\linewidth]{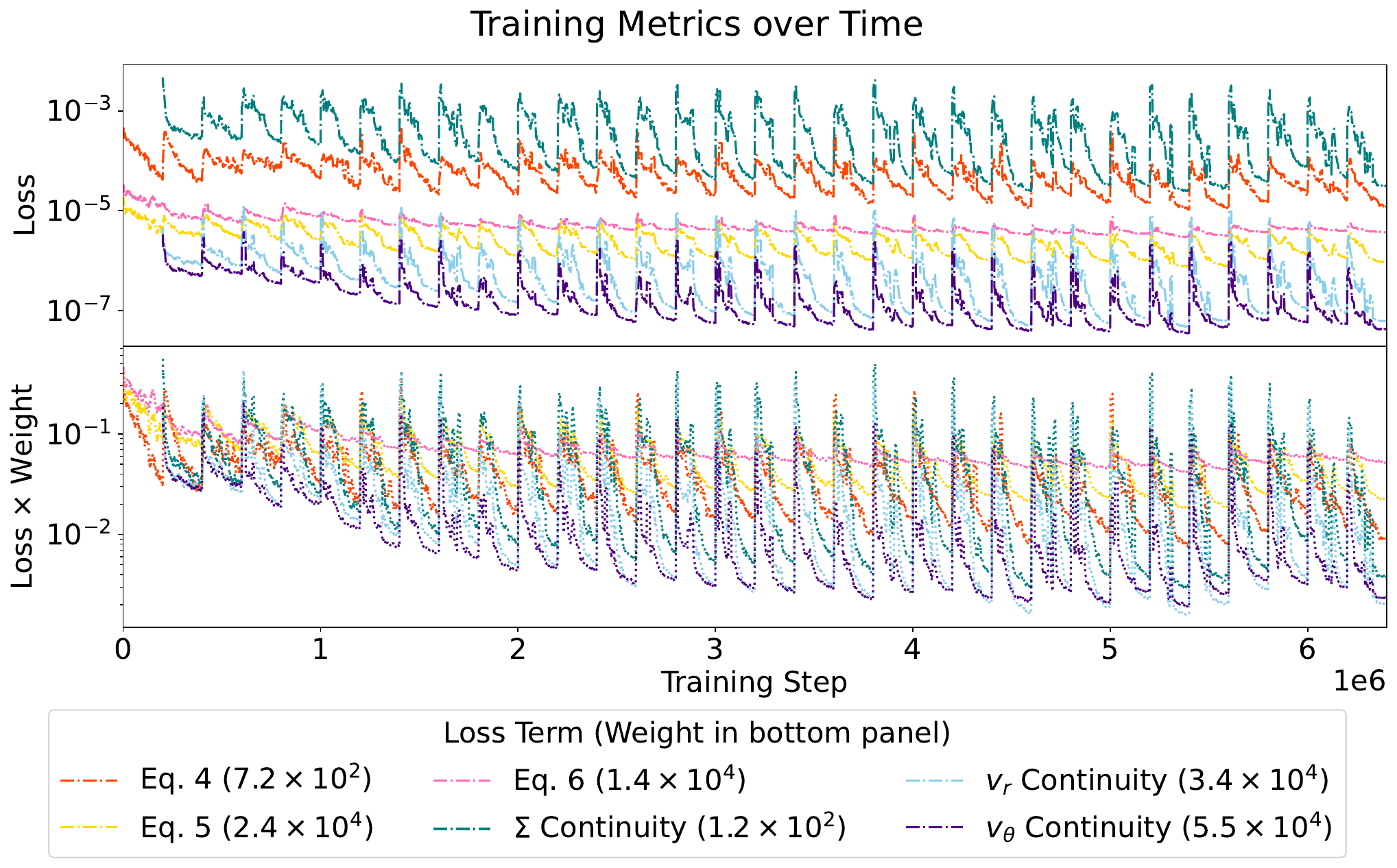}
    \caption{
    Training loss history for the fiducial model ($M_{\rm 2}/M_{\rm 1}=10^{-3}$) presented in \S\ref{sec:solutions_for_q1e-3}. The top panel shows the evolution of the unweighted mean squared error (MSE) for each component of the loss function, including PDE residuals for mass continuity (Eq.~\ref{eq:PDE1_inertial}), radial (Eq.~\ref{eq:PDE2_rotating}) and azimuthal (Eq.~\ref{eq:PDE3_rotating}) components of Navier-Stokes equations, alongside data-matching losses that enforce continuity across time-marching windows. The bottom panel shows the same components after being scaled by their respective weights (\S\ref{sec:competing_loss_terms_and_weight_balancing}), representing their actual contribution to the total loss during optimization. 
    }
    \label{fig:training_loss}
\end{figure*}

Figure~\ref{fig:training_loss} illustrates our model's training dynamics. The top panel shows the loss components over the course of training. These components include the mean squared PDE residuals, alongside the mean squared continuity errors for surface density ($\Sigma$), radial velocity ($v_r$), and azimuthal velocity ($v_\theta$) between adjacent time-marching windows.
As training proceeds sequentially through 32 time-marching windows (see \S\ref{sec:time_marching}), the loss curves exhibit periodic spikes that correspond to the random re-initialization of the sub-network for each new window. The top panel reveals that the raw PDE and data loss components differ by several orders of magnitude. This imbalance poses a significant challenge, as components with larger magnitudes can dominate the optimization and hinder convergence. To mitigate this, we multiply fixed weights to each loss component (\S\ref{sec:competing_loss_terms_and_weight_balancing}).
The bottom panel demonstrates the effectiveness of this strategy, showing that the rescaled loss components are comparable in magnitude. This balance is essential for stable and accurate solutions, as it ensures all PDE and continuity components contribute meaningfully to the optimization process.

While this training process successfully converges to a solution, it incurs a significant computational cost compared to traditional solvers. A FARGO3D simulation of 10 orbits completes in three minutes on a single NVIDIA A100 GPU, whereas our PINN requires $\approx50$ hours on the same hardware.
This performance gap stems from three primary factors: the intensive computation required for evaluating partial derivatives in the PDEs, slow network convergence necessitating 200,000 iterations per sub-network, and the sequential training of 32 time-marching sub-networks.

\subsection{Disk-companion system parameter variation tests} \label{sec:performance_under_different_disk_parameters}
\begin{figure*}
    \centering
    \includegraphics[width=0.8\linewidth]{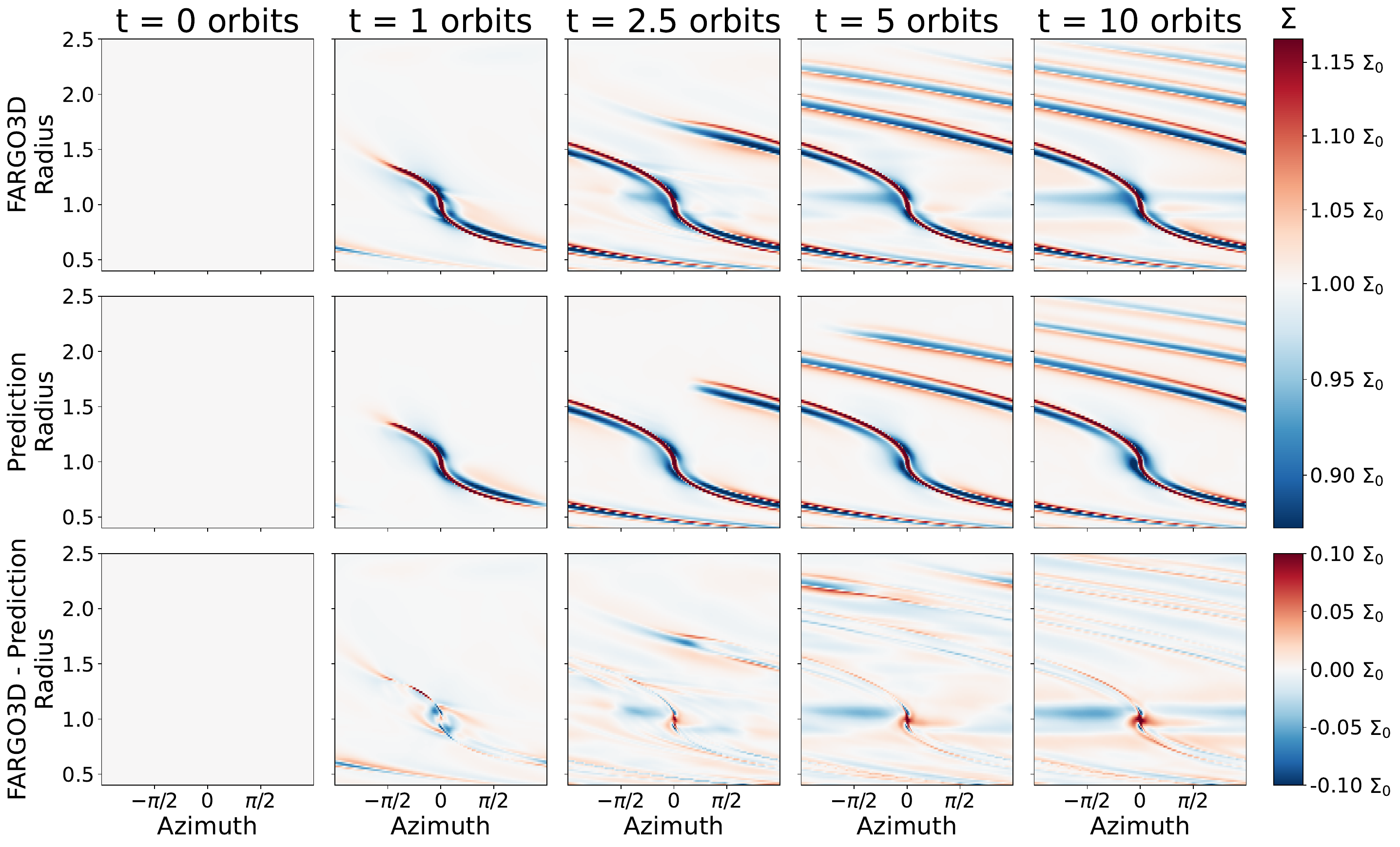}
    \caption{
    Same as Figure~\ref{fig:pred_fargo_compare_sigma}, but for the case with a lower-mass secondary ($M_{\rm 2}/M_{\rm 1}=10^{-4}$; \S\ref{sec:performance_under_different_disk_parameters}).
    The secondary induces weaker density perturbations compared to the fiducial $M_{\rm 2}/M_{\rm 1}=10^{-3}$ case. An animation of this figure is available at \url{https://doi.org/10.6084/m9.figshare.30192904}.}
    \label{fig:pred_fargo_compare_sigma_q1e-4}
\end{figure*}

We test our PINN's robustness by varying two system parameters. First, for a lower mass with $M_{\rm 2}/M_{\rm 1}=10^{-4}$ (Figure \ref{fig:pred_fargo_compare_sigma_q1e-4}), the model captures the weaker spiral wave pattern but fails to reproduce the shallow density gap. In this scenario, the model struggles to accurately resolve the sharp density depletion near the secondary's orbit.

\begin{figure*}
    \centering
    \includegraphics[width=0.8\linewidth]{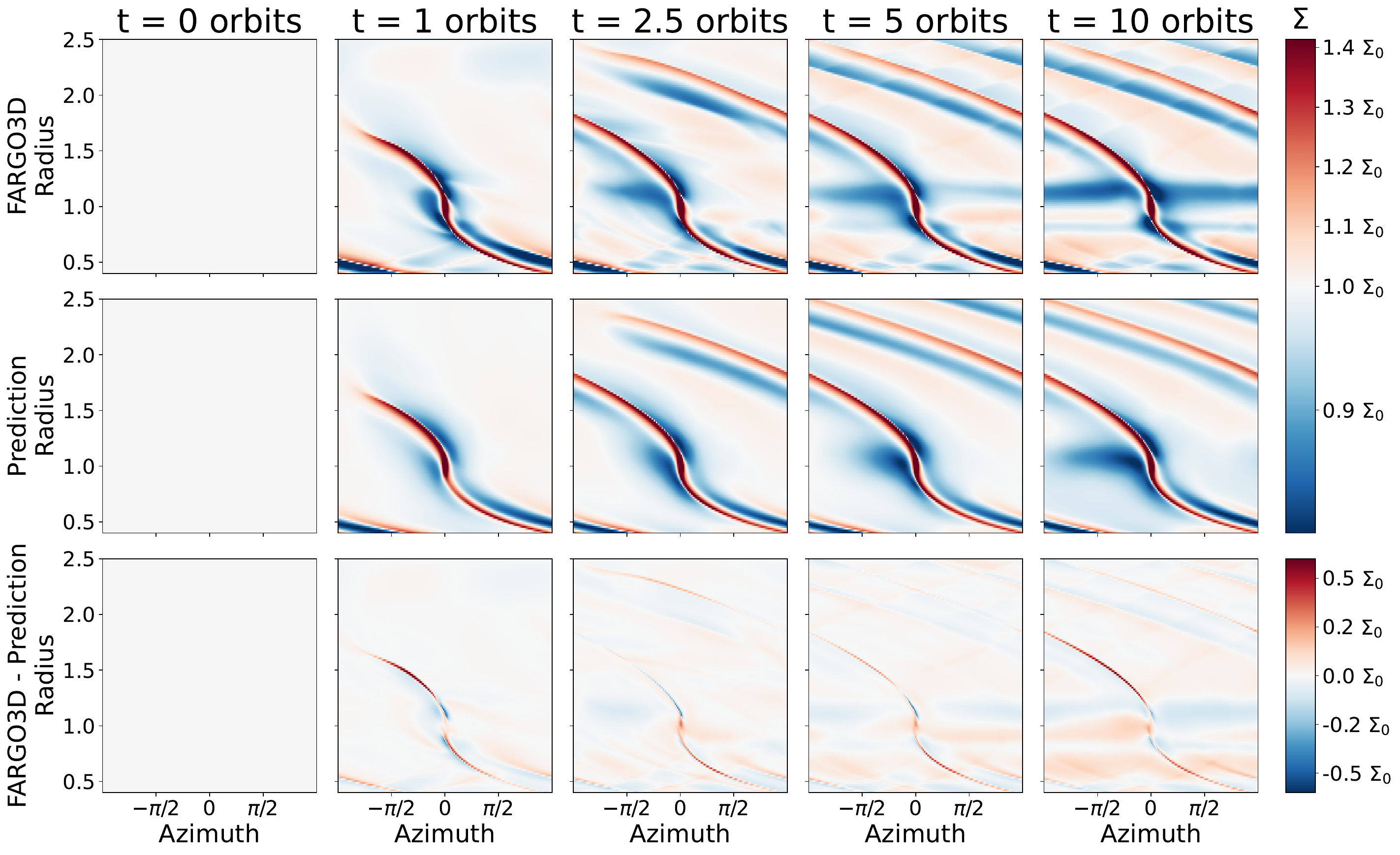}
    \caption{
    Same as Figure~\ref{fig:pred_fargo_compare_sigma}, but for the case with a higher disk aspect ratio $h=0.1$(\S\ref{sec:performance_under_different_disk_parameters}).
    While the PINN recovers the main spiral features, it fails to reproduce the deep density gap near the secondary's orbit. An animation of this figure is available at \url{https://doi.org/10.6084/m9.figshare.30192904}.
    }
    \label{fig:pred_fargo_compare_sigma_q1e-3_ar01}
\end{figure*}

\begin{figure*}
    \centering
    \includegraphics[width=0.8\linewidth]{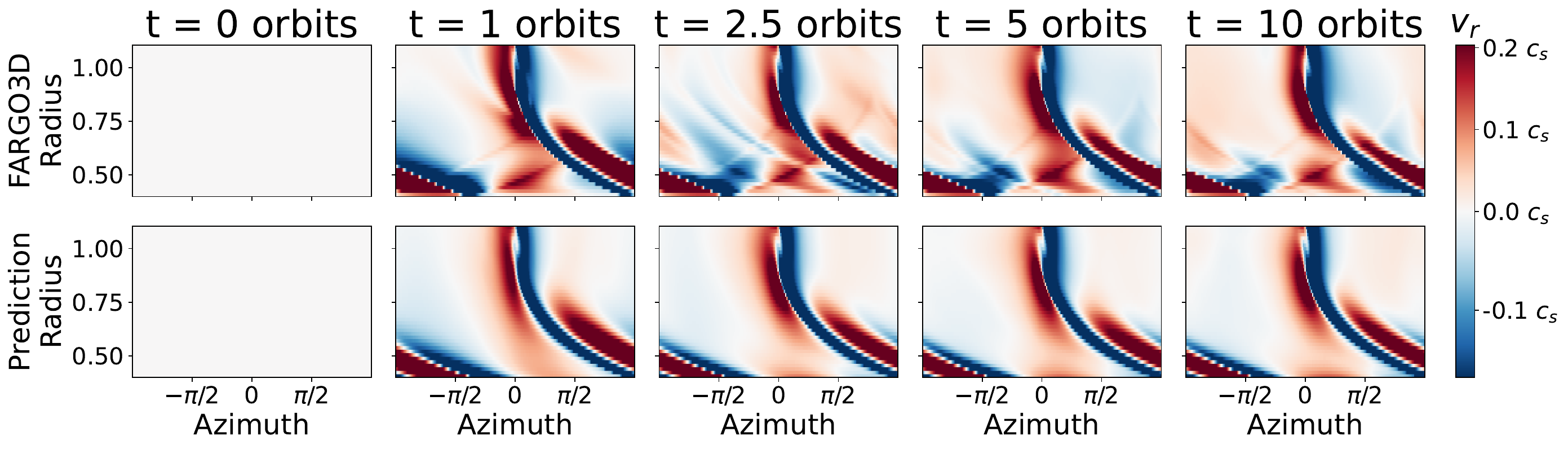}
    \caption{
    Comparison of radial velocity fields from FARGO3D and PINN for a secondary with $M_{\rm 2}/M_{\rm 1}=10^{-3}$ in a disk with aspect ratio $h=0.1$ in \S\ref{sec:performance_under_different_disk_parameters}. The PINN solution shows a smoother velocity field near the inner boundary ($r=0.4 r_{\rm 2}$), while FARGO3D exhibits artificial wave reflections despite damping conditions.}
    \label{fig:pred_fargo_compare_vr_q1e-3_ar01}
\end{figure*}

In the higher–aspect-ratio case ($h=0.1$; Figures~\ref{fig:pred_fargo_compare_sigma_q1e-3_ar01} and \ref{fig:pred_fargo_compare_vr_q1e-3_ar01}), the PINN reproduces the spiral structure but not the deep density gap seen in FARGO3D. However, it avoids the reflected velocity perturbations present near the inner boundary ($0.4r_{\rm 2} < r < r_{\rm 2}$) in FARGO3D, owing to its intrinsic bias toward smooth function extrapolation, which naturally handles open boundaries without artificial wave-damping.

\section{Summary and Discussion}\label{sec:conclusion}
We present a comprehensive evaluation of physics-informed neural networks (PINNs) for solving the two-dimensional, time-dependent compressible Navier-Stokes equations governing thin accretion disks without labeled solution data. Our approach embeds the governing equations directly into the loss function and integrates several advanced PINN techniques: time-marching for long-term evolution (\S\ref{sec:time_marching}), natural wave-killing boundary conditions (\S\ref{sec:radial_boundary_conditions}), periodic coordinate transformation (\S\ref{sec:azimuthal_boundary_conditions}), output scaling (\S\ref{sec:initialization_bias}), hard constraints for initial conditions (\S\ref{sec:initial_condition_and_hard_constraint}), and loss balancing (\S\ref{sec:competing_loss_terms_and_weight_balancing}).

Our investigation highlights three primary achievements of this framework.
\begin{itemize}
    \item The PINN successfully reproduces spiral density waves, gap formation, and associated velocity fields across a range of parameter settings when validated against FARGO3D simulations.
    \item The boundary-free approach eliminates spurious wave reflections at radial edges, providing a significant advantage over traditional numerical methods that require carefully designed damping zones and absorbing boundary conditions.
    \item The PINN's continuous functional representation enables efficient evaluation at arbitrary spatiotemporal points without interpolation, offering computational flexibility that traditional grid-based methods find difficult to match.
\end{itemize}

However, the current approach has limitations that warrant further investigation:
\begin{itemize}
    \item Fine-scale flow structures near the secondary and gap edges appear smoothed compared to the baseline finite-difference solution from FARGO3D, indicating that the current training setup under-resolves sharp gradients.
    \item The computational cost remains substantial—training requires $\approx50$ hours on an A100 GPU, three orders of magnitude longer than a FARGO3D simulation on the same hardware.
\end{itemize}

Despite these limitations, the present work shows that PINNs can already reproduce the essential physics of accretion disk systems in a data-free setting and offer a streamlined alternative to traditional numerical solvers.

Future research will improve the current model and training strategy. Recent advances in PINNs on benchmark problems claim to reduce training costs with lightweight architectures \citep{cho2023separable}, boost solution accuracy with more stable and efficient architectures \citep{wang2024piratenets,cao2025analysis,liu2025bwler,wu2025propinn,wang2025fundiff}, or improve optimization stability \citep{cao2023tsonn,cao2024solver,wang2025gradient}.
With these improvements, the framework can be applied to longer time evolution toward steady states and extended to more demanding astrophysical scenarios, including fully three-dimensional disks and multi-planet systems. Additionally, developing a parameterized PINN by including physical parameters (e.g., secondary mass) as network inputs can create a powerful surrogate model for large-scale parameter studies \citep{wang2021learning}.

\section*{Acknowledgements}
We are grateful to an anonymous referee for constructive suggestions that improved our paper.
We thank Jaehan Bae, Xuening Bai, Pablo Ben{\'\i}tez-Llambay, Shengze Cai, Shengtai Li, Miles Cranmer, Bin Dong, Scott Field, Jeffrey Fung, Xiaotian Gao, Ka Wai Ho, Jiequn Han, Pinghui Huang, Pengzhan Jin, Xiaowei Jin, Hui Li, Tie-Yan Liu, Zhiping Mao, Chris Ormel, Wenlei Shi, Karun Thanjavur, Yiwei Wang, Yinhao Wu, Zhenghao Xu, Ka Ho Yuen, Minhao Zhang, Wei Zhu for help and useful discussions in the project.

S.M. and A.I. acknowledge support from the National Aeronautics and Space Administration under grant No. 80NSSC18K0828. S.M., W.W. and R.D. are supported by the Government of Canada's New Frontiers in Research Fund (NFRF), [NFRFE-2022-00159].
This research was enabled in part by support provided by the Digital Research Alliance of Canada (\url{alliance.can.ca}), the National Energy Research Scientific Computing Center (NERSC) under award number FES-ERCAP-m4239, and by the High-performance Computing Platform of Peking University.

%\clearpage

%\appendix

\bibliography{sample631}{}
\bibliographystyle{aasjournal}



\end{document}